\documentclass[reprint,superscriptaddress,pra,floatfix]{revtex4-2}
\usepackage{amsmath}
\usepackage{graphicx}
\usepackage{natbib}
\graphicspath{{figures/}}
\begin{document}
	\title{Investigation on trapping capability of circular swallowtail beams}
	\author{Huanpeng Liang}
	\author{Xiaofang Lu}
	\author{Kaijian Chen}
	\author{Haixia Wu}
	\author{Nana Liu}
	\author{Liu Tan}
	\author{Shaozhou Jiang}
	\author{Yi Liang}
	\email{liangyi@gxu.edu.cn}
	\affiliation{Guangxi Key Lab for Relativistic Astrophysics, Center on Nanoenergy Research, Guangxi Colleges and Universities Key Laboratory of Blue Energy and Systems Integration, School of Physical Science and Technology, Guangxi University, Nanning, Guangxi 530004, China}
	
	\date{\today}
	\begin{abstract}
Circular swallowtail beams (CSBs) with their remarkable autofocusing capability have garnered significant interests due to their potential applications in optical trapping. This study delves into a comprehensive investigation of the trapping force properties of CSBs. Through a combination of experimental observations and theoretical analysis, we systematically explore the quantitative manipulation of trapping forces by adjusting specific parameters. This detailed investigation provides insights into the trapping force performance and stability of CSBs. Furthermore, the experimental validation of particle trapping using CSBs underscores their effectiveness, emphasizing their significant potential for optical manipulation and trapping applications. 
	\end{abstract}
	
	\maketitle
	
	\section{Introduction}
Various accelerating beams that propagate along curved trajectories have attracted a lot of attention due to a huge potential for many applications including optical tweezers\cite{2021OTreview_Chen,2021trapreview_Yang,1986Ashkin}, biomedical science\cite{2020OTreview_Campugan,RN316}, astrophysics\cite{2021OTastronomy}and so on. The most famous accelerating beams, named Airy beams\cite{2007AiryAccelerating,2010CABRayleigh,2011Airytrap_Zheng,2011Airytrap_Kaminer,2008Selfhealing,2011Airytrap_Zhang}, are experimentally generated in 2007. Airy beams can be described as a result of the fold catastrophe in catastrophe theory. There are seven fundamental catastrophes including fold\cite{2007AiryAccelerating}, cusp\cite{2012CPBselfhealing}, swallowtail\cite{2017Swallowtail&Butterfly}, butterfly\cite{2017Swallowtail&Butterfly}, elliptical umbilic\cite{RN330}, hyperbolic umbilic\cite{RN329}, and parabolic umbilic\cite{RN319}. Most beams related to these catastrophes can present a curve propagation trajectory, i.e., and therefore they can be employed to generate an accelerating beam. For example, a family of accelerating beams, called Pearcey beams\cite{2012CPBselfhealing,2018CPB_Chen}, comes from the cusp catastrophe. Recently, originating from swallowtail catastrophe, a new kind of high-order accelerating beams named Swallowtail beams are generated\cite{2017SwallowtailDynamics}. Swallowtail beams exhibit a unique feature: by tuning control parameters, Swallowtail beams not only can evolve into higher-order butterfly catastrophes during propagation, but also can decay to lower-order cusp catastrophe such as Pearcey beams. 
    
 As we know, the autofocusing properties of autofocusing beams\cite{2007AiryGauss,2007AiryObservation,2008Airypoynting,2011Airy_Granot,2019Airyreview_Efremidis}generated from accelerating beams is determined by the acceleration of the beams. The autofocusing propagation of circular Airy beams (CABs) can be modulated by tuning the parameters related to acceleration of Airy beams including the scale factor and initial position of main lobe\cite{2013CABRayleigh_Jiang}. Along this line, one can infer that autofocusing beams based on Swallowtail beams such as Circular swallowtail beams (CSBs) also show a similar autofocusing property, as demonstrated in Refs.\cite{2017CSBpoynting_Cheng,2017Swallowtail&Butterfly,2021CSB_TengHou,2021swallowtail_TengHou,2022CSBtornado,2022BSwallowtailottle,2022Swallowtailpolycyclic}. Due to this autofocusing property, CSBs exhibits a stronger intensity contrast at the autofocusing position, which means that it has great potential for optical manipulation or trapping. However, as far as we know, trapping force and the trapping stability of CSBs lacks systematic study, especially in experiment. 
 
 Thus, in this paper, we conduct a comprehensive investigation on trapping force and stability of circular swallowtail beams. Initially, we delve into the autofocusing characteristics of circular swallowtail beams through both experimental and theoretical approaches. Subsequently, we extend our analysis to the trapping performance on Rayleigh particles, providing theoretical insights into the trapping forces exerted by CSBs. To validate our findings, we employ CSBs as optical tweezers in experimental, trapping particles and measuring the optical trap stiffness. Remarkably, our measured values agree well with those calculated using Generalized Lorentzian Mie theory, confirming the accuracy and reliability of CSBs in optical trapping applications. This work significantly contributes to the understanding of the optical trapping capabilities of CSBs, paving the way for the application of these innovative tools in optical manipulation and trapping scenarios.

	
	\section{Propagation of CSBs}
	Based on catastrophe theory, a caustic field with standard diffraction integral is given by\cite{2017Swallowtail&Butterfly}:
	\begin{equation}
		\psi_{n}(\textbf{a})=\int_{-\infty}^{+\infty}exp[ip_n(\textbf{a},s)]ds \label{eq1}
	\end{equation}
	where $p_n(\textbf{a},s)$ is the canonical potential function that can determine the properties of the caustic filed, and it is defined by\cite{2017Swallowtail&Butterfly}:
	\begin{equation}
		p_n(\textbf{a},s)=s^n+\sum_{j=1}^{n-2}a_js^j \label{eq2}
	\end{equation}
	Here, $\textbf{a}=\left({a}_{1},{a}_{2},\dots,{a}_{j}\right)$ is the control parameter and $s$ is the state parameter. Usually, a Swallowtail beam can be expressed by the swallowtail catastrophe integral\cite{2021CSB_TengHou}:
	\begin{equation}
		Sw(X,Y,Z)=\int_{-\infty}^{+\infty}exp([i(s^5+Zs^3+Ys^2+Xs^1)])ds \label{eq3}
	\end{equation}
	where  $X,Y,Z$ denotes the dimensionless coordinates in the real space.
	
	For simply constructing a circular swallowtail beam, Eq.\ref{eq3} is transformed into a expression in cylindrical coordinates\cite{2021CSB_TengHou}:
	\begin{equation}
		\Phi(r,\theta,0)=Sw\bigg(\frac{r_0-r}{w_0},0,0\bigg)Q(r,\theta) \label{eq4}
	\end{equation}
	where $r_0$ is the initial radius of the main ring and $w_0$ is the scale factor, which determines the numbers and width of ring. $Q(r,\theta)$ is the optical aperture to make the energy of beam finite for experimental realization. It restricts the beams within given region and  is defined as follows:
	\begin{equation}
		Q(r,\theta)=
		\begin{cases}
			1, \quad  0 \le r \le R_B, 0\le \theta \le 2\pi
			\\ 0, \quad other. 
		\end{cases}
		\label{eq5}
	\end{equation}
	where $R_B$ is the radius of the CSBs in the initial plane. 
	
\begin{figure}[htbp]
		\centering
		\includegraphics[width=\linewidth]{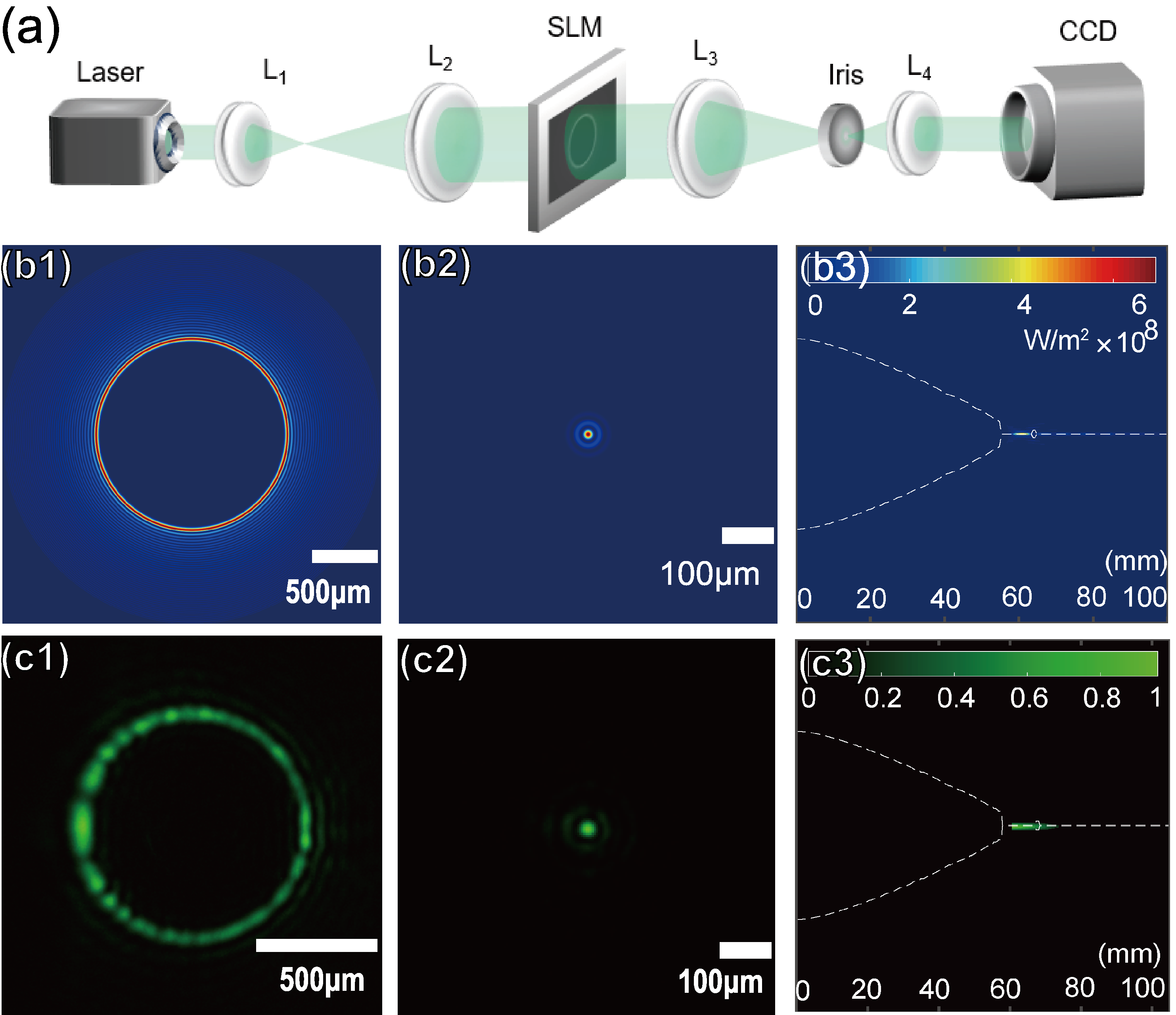}
		\caption{Propagation of CSBs. (a) Experiment setup: Laser(532 nm), semiconductor laser. L, lens. SLM, spatial light modulator. (b1, b2) Numerical results at z=0 mm,61 mm. (b3) Numerical propagation sideview. (c1-c3) corresponding experimental results.}
		\label{fig1}
	\end{figure}

	In paraxial approximation, Swallowtail beam propagates following the below wave equation\cite{2021CSB_TengHou}:
	\begin{equation}
		2i\frac{\partial \Phi}{\partial \xi}
		+\frac{\partial^{2} \Phi}{\partial r^{2}}
		+r^{-1} \frac{\partial \Phi}{\partial r}
		+r^{-2} \frac{\partial \Phi}{\partial \theta}
		=0
		\label{eq6}
	\end{equation}	
	Then, similar to the previous works\cite{2021Airy_FuxiLu,2021BesselCAB_FuxiLu}. We can apply beam propagation methods to simulate the propagation of CSBs and analyze their autofocusing property. In experiment, CSBs are generated by means of hologram, as shown in the Fig.\ref{fig1}(a). The hologram is loaded in a transmissive spatial light modulator(SLM) (1024$\times$768, pixel pitch is 36 $\mu m$, fill factor is $58\%$). A linear polarized Gaussian beam is emitted from the semiconductor laser (532 nm), and then are expanded by the lens (L1 and L2). After passing though the SLM, such beam is transferred into a CSB with sufficient information by a 4f system (L3 and L4). The initial intensity distribution of the CSBs appears in the focal plane of L4 and the propagation of CSBs can be recorded by moving the CCD camera. In the simulation and experiment, the parameters are set as $R_B$=1.4 mm, $r_0$=0.7 mm, $w_0$=8.33 $\mu m$, the incident power of simulation is kept as 1 W. 

     Figure\ref{fig1}(b, c) display the numerical and experimental propagation results of CSBs including the intensity of propagation of CSBs at z=0 mm,61 mm (z is the propagation distance) and their propagation sideview. Clearly, at the beginning, the main ring of CSBs possesses the maximum intensity when CSBs are generated in Fig.\ref{fig1}(b1). When the beam comes to the position of z=61 mm [Fig.\ref{fig1}(b2)], all power automatically focuses at a very small spot whose radius is about $10.5\enspace \mu m$, obviously exhibiting a autofocusing property without any lens in free space. 
	Fig.\ref{fig1}(b3,c3) are the sideview of propagation, the dash lines depict the trajectory of the main ring and autofocusing is observed clearly. One can find that the experiment are basically consistent with the simulation. 
	
	\section{Trapping performance of CSBs on Rayleigh particle}
	According to the theory, the gradient force $\vec{F_g}$ and the scattering force $\vec{F_s}$ of Rayleigh particle can be calculated by\cite{RN324}:
	\begin{equation}
		\vec{F_g}=\frac{1}{4} \varepsilon_0 \varepsilon_m Re(\alpha) \nabla \left| \vec{\Phi}^2 \right|
		\label{eq7}
	\end{equation}
	\begin{equation}
		\vec{F_s}=\frac{1}{6\pi c} \epsilon^3_m k^4_0 \left| \alpha^2 \right| \vec{S}
		\label{eq8}
	\end{equation}
	where $\epsilon_m$ is the dielectric of the medium around the particle, $\epsilon_0$ is the dielectric of the constant in vacuum, $k_0$ is the wave number,  $\alpha=4\pi R_P^3({\varepsilon_p-\varepsilon_m})/({\varepsilon_p+2\varepsilon_m})$  is polarizability ( $R_P$ is the radius of Rayleigh particles). $\vec{S}$ is the Poynting vector, which can be cauculated by\cite{2008Selfhealing,RN325}:
	\begin{equation}
		\vec{S}=\vec{S_z}+\vec{S}_{\perp}=
		\frac{1}{2\eta_0}\left| \Phi^2 \right| \hat{z} +
		\frac{i}{4\eta_0k} \left[ \Phi \nabla_\perp \Phi^\ast - \Phi^\ast \nabla_\perp \Phi \right]
		\label{eq10}
	\end{equation}	
	where the $\eta_0=\sqrt{\mu_0\ / \varepsilon_0}$ is the impedance of free space. 
	
	\begin{figure}[htbp]
		\centering
		\includegraphics[width=\linewidth]{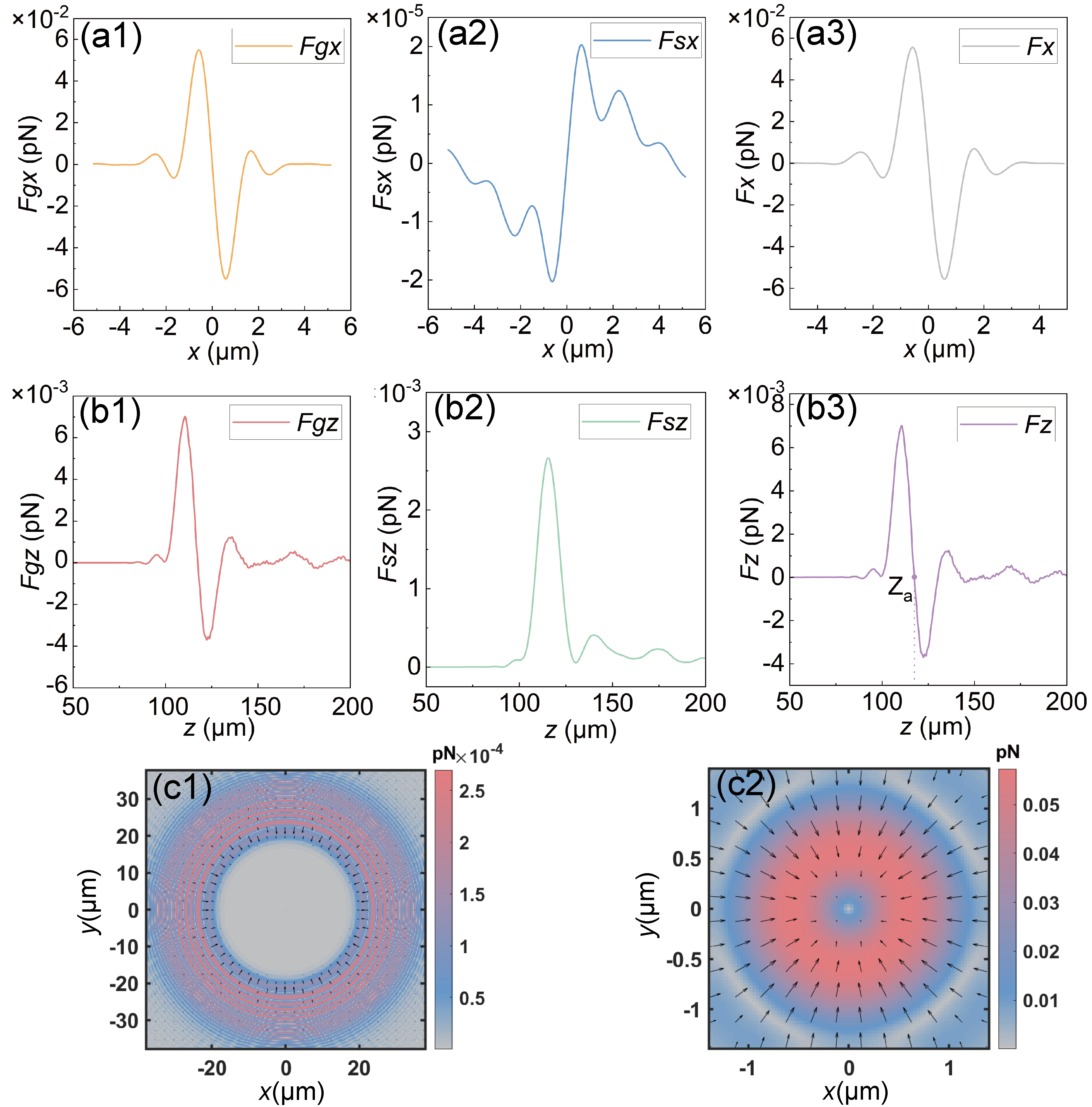}
		\caption{Distribution of tapping force. The $R_B$ is 50 $\mu$m, the $r_0$ is 25 $\mu$m, the $w_0$ is 0.7 $\mu$m, the radius of polystyrene particles is 46 nm. The focal length is 116.29 $\mu$m. $Z_a$ = 117.29 $\mu$m is the longitudinal trapping position. (a1) Transversal gradient force ($F_{gx}$) at $Z_a$. (a2) Transversal scattering force ($F_{sx}$) at $Z_a$. (a3) Transversal total force ($F_{x}$) at $Z_a$. (b1) Longitudinal gradient force ($F_{gz}$). b(2) Longitudinal scattering force ($F_{sz}$). (b3) Total Longitudinal force ($F_z$). (c1) Force distribution at  z=30 $\mu$m. (c2) Force distribution at the focus.}
		\label{Fig2}
	\end{figure}
	
	Figure \ref{Fig2} presents the calculated trapping force results of CSBs with a  focal spot of 2.76 $\mu$m while they trap Rayleigh polystyrene particles of $R_P$ = 46 nm in water ($\varepsilon_m=1.77$, $\varepsilon_p=2.47$). As illustrated in Figs.\ref{Fig2}a(1, 2) and b(1, 2), both of transversal and longitudinal gradient forces are larger than the scattering force at the focus. Furthermore, we find that the maximum value of the total axial force exceeds than the difference between buoyancy and gravity (as calculated by us at $2\times10^{-10}$ pN, and $10^{-7}$ pN according to Ref.\cite{2013CABRayleigh_Jiang}). In this case, CSBs is possible to trap the above particles stably. Noted that, all force vectors are directed towards the center of the beams, indicating that there is only one transverse trapping position and all surrounding particles are pulled towards the center [Fig.\ref{Fig2}c(1,2)].   
 
     To further investigate the trapping stability of CSBs, we calculate the related parameters including the Brownian force $F_b$ of  trapped particle and the Boltzmann factor as follows\cite{1986Ashkin,2013CABRayleigh_Jiang,2019Brownmotionformular}:
    \begin{equation}
		F_b=\sqrt{12{\pi}{\eta}R_pk_BT}
		\label{eq13}
    \end{equation}
     \begin{equation}
		\xi=exp(-U/k_BT)
		\label{eq11}
   \end{equation}
		where $U=\frac{1}{4}\varepsilon_0 \varepsilon_m Re(\alpha) \Delta \left|\vec{\phi}^2\right|$ is the potential energy of gradient force,  $\Delta
	\lvert \vec{\Phi}^2 \rvert$ denotes the intensity difference related to the potential energy of the gradient force, $k_B$ represents the Boltzmann constant, $T$ is the thermodynamic temperature of medium (here, we assume that the $T=300\enspace K$), $\eta$ denotes the viscosity of the surrounding medium(Water, $\eta=8\times10^{-4}\enspace Pa\cdot s$).   
	
	\begin{table}[htbp]
		\centering
		\caption{Trapping stability analysis.}
		\label{table1}
		\begin{ruledtabular}
		\begin{tabular}{ccccc}
		$F_b$(pN) & $F_{gx}$(pN)  & $F_{gz}$(pN) & $\xi$ \\
			\colrule
			$2.40\times10^{-3}$ & $5.32\times10^{-2}$ & $5.26\times10^{-3}$ & $1.94\times10^{-5}$ \\
		\end{tabular}
		\end{ruledtabular}
	\end{table}
		The  calculation results related to the trapping stability are presented in Table.\ref{table1}. Clearly, both the transversal gradient force ($F_{gx}$) and longitudinal gradient forces ($F_{gz}$) can overcome the Brownian force ($F_b$).  Moreover, the Boltzmann factor ($\xi$) is much less than 1, suggesting that the time to trap a particle is much less than the time to leave the trap due to Brownian motion\cite{1986Ashkin}. Thus, it can be inferred that the above Rayleigh particles can be stably trapped by CSBs. 
 
	\section{Tuning the autofocusing and trapping force properties}
	\begin{figure}[htbp]
		\centering
		\includegraphics[width=\linewidth]{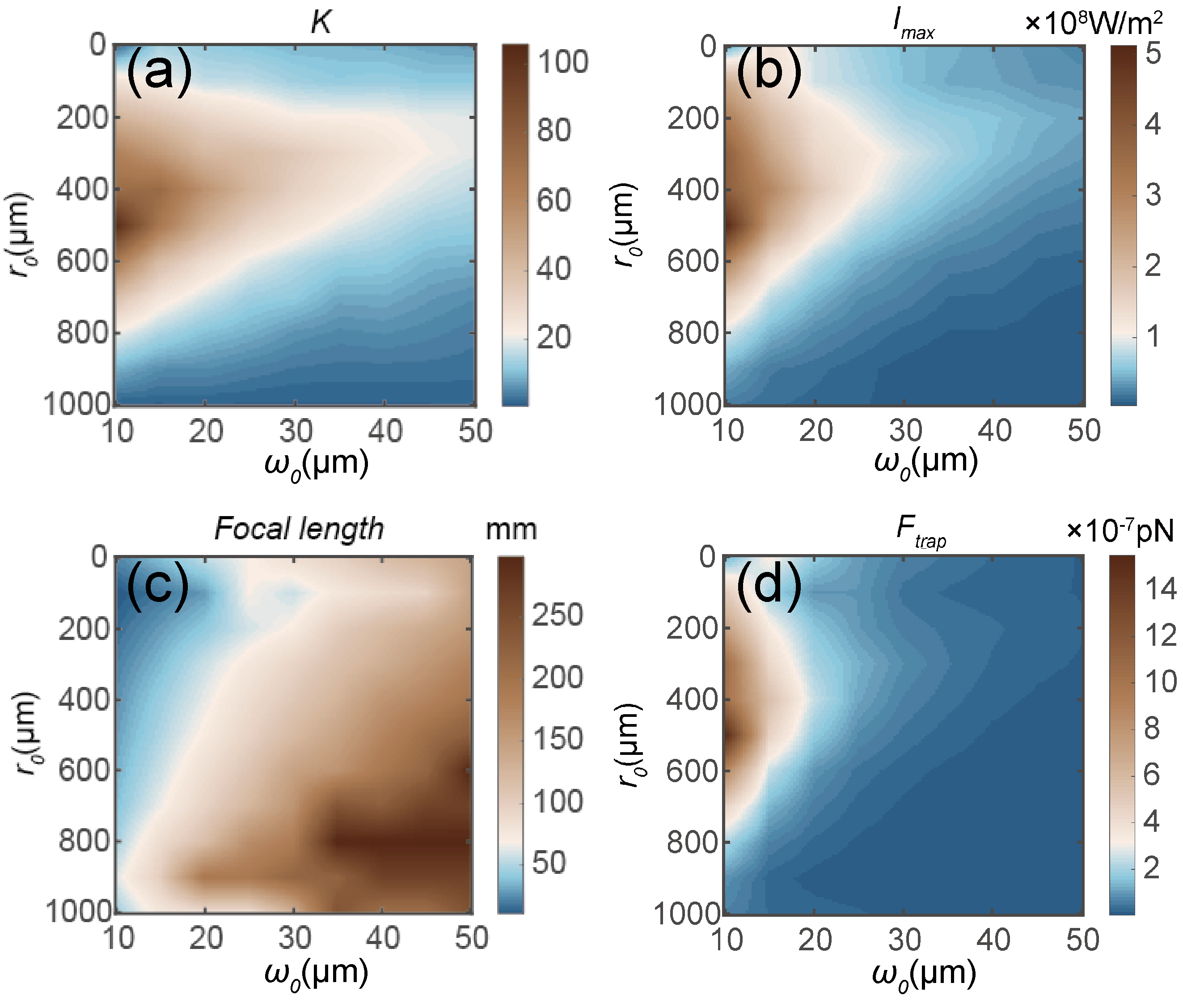} 
		\caption{Trapping-force performance of CSBs with varying parameters $r_0$ and $w_0$. $R_B$ is set to be 2 mm, the incident power is 1 W, the wavelength is 1064 nm, the radius of polystyrene particle is 20 nm. (a) The Autofocusing properties, $K=I_{max}/I_0$, $I_{max}$ is the focal peak intensity, $I_0$ is the peak intensity of initial plane. (b) The focal peak intensity $I_{max}$. (c) The autofocusing length z. (d) The trapping-force $F_{trap}$.}
		\label{Fig3}
	\end{figure}
	
	As we know, the $r_0$ controls the radius of main ring in the source plane, and the $w_0$ controls the width of the main ring and the number of rings in the source plane. Both of them affect the focal spot size and further the autofocusing intensity. Thus, by altering the parameters of CSBs in detail, we can analyze the quantitative changes in autofocusing and trapping force at the focus, as shown in Fig.\ref{Fig3}. From Fig.\ref{Fig3}, one can see that, the distribution of focal peak intensity ($I_{max}$) in Fig.\ref{Fig3}(b) is different from the intensity contrast $K$ ($K$ is the ratio between the maximum light intensity at the focal plane and the maximum light intensity at
 the initial plane) [Fig.\ref{Fig3}(a)]. However, both of focal peak intensity and intensity contrast can reach at a largest value when choosing appropriate $r_0$ while they decreases with increasing $w_0$. During these changing, the autofocusing length keeps increaseing and the trapping force following the variation of focal peak intensity [Fig.\ref{Fig3}(c ,d)]. Especially, it is found that the theoretical optimal autofocusing and largest trapping force appears at $r_0\approx0.5R_B$ [Fig.\ref{Fig3}(d)]. Because the variation of the trapping force is larger than that of intensity contrast at the same conditions, trapping force is more sensitive to the influence of the parameters.  
 
	\section{Trapping performance of CSBs on Mie particles}
	\begin{figure}[htbp]
		\centering
		\includegraphics[width=\linewidth]{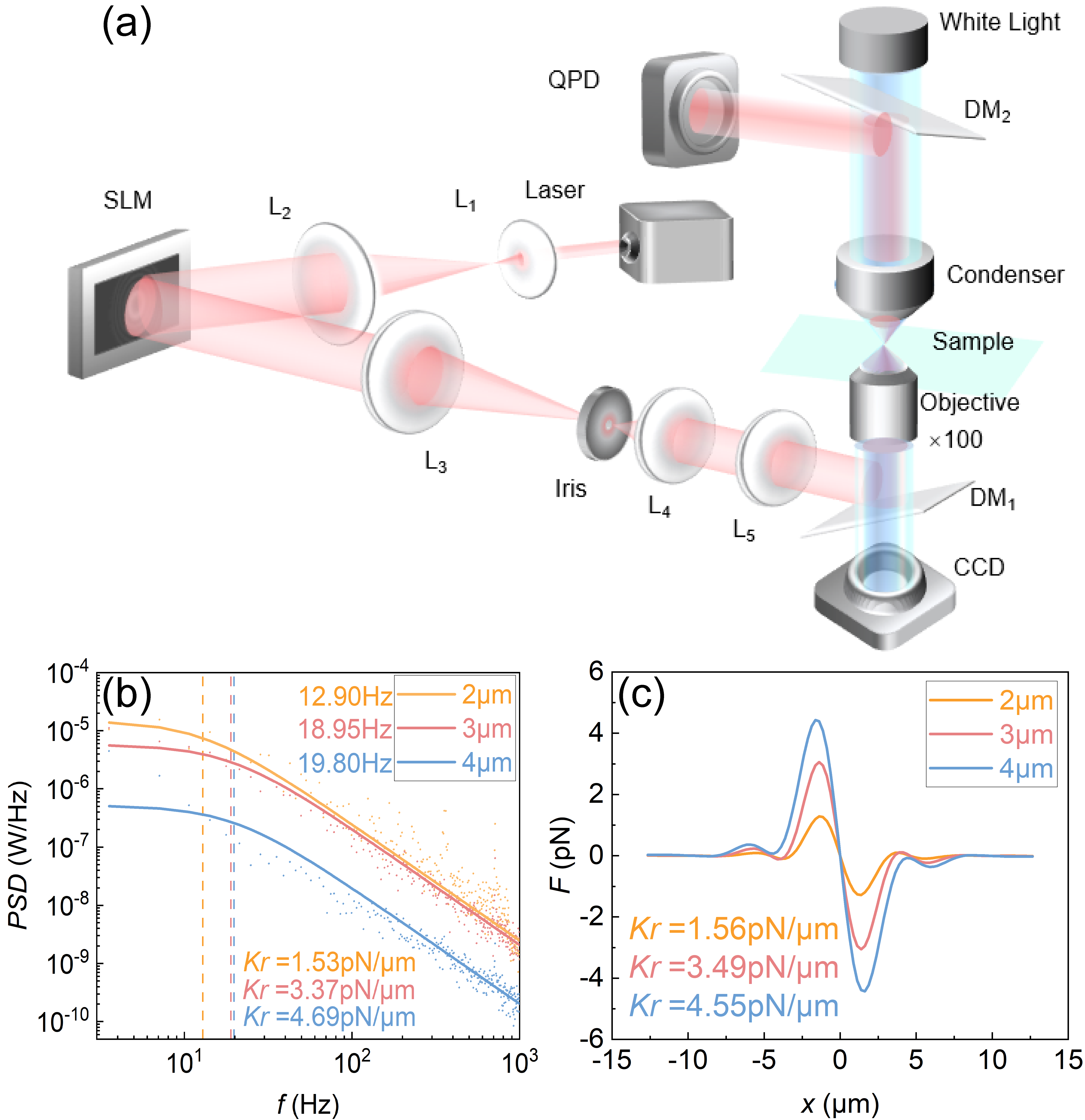}
		\caption{The experiment setup for trapping and observing particles and the power spectra and trapping stiffness. Laser, wavelength is 1064 nm. SLM, Spatial Light Modulator. CCD, Charged Coupled Device. Objective, oil objective lens($\times100$), the numerical aperture is 1.25. DM, Dichroic Mirror. QPD, quadrant photoelectric detector. Incident power is 20 mW. (b) Power spectra of trapping polystyrene particles with a diameter of 2-4 $\mu m$. (c) The trapping force distribution of Mie particles, and the trap stiffness is given.}
		\label{fig4}
	\end{figure}	 
 	
	Finally, to investigate the trapping performance of CSBs on Mie articles, we utilized the  CSBs as optical tweezers to trap large polystyrene beads with different sizes in water (See video within the Supplemental Material \cite{video}). As shown in [Fig.\ref{fig4}(a)], we employed a similar experimental setup as in Refs. \cite{2021BesselCAB_FuxiLu,RN324,RN325}: The CSB was generated at the focal plane of Lens 4 (source plane of the CSB), and subsequently relayed to the sample by using a 4-f imaging system comprised of Lens 5 and Oil Lens. After changing the distances between Len4 and Lens5, we placed the autofocusing position right at the focal point of Oil Lens. This allowed us to realize optical tweezers based on CSBs for trapping polystyrene beads. We then used power spectrum methods \cite{ 2021BesselCAB_FuxiLu,RN324,RN325} to analyze the trap stiffnesses of the beams. 
 
In detail, to measure the trap stiffness, we collected scattered light from beads using a condenser lens and a quadrant photodiode (QPD) to record the real-time positions of the trapped beads. These positions were then transformed into a power spectrum for calculating the trapping stiffness. Specifically, the trap stiffness ($\kappa_r=-dF_{trap}/dr$)
 of the beams was calculated from the experimental data of the real-time trapped bead positions using the Langevin equation and the corner frequency power spectrum ($f_{c,r}$): 
 $f_{c,r}=\kappa_r/2\pi\gamma$. Here, $\gamma$ represents the particle friction coefficient 
 ($\gamma=3\pi\eta D_P$), where $\eta$ is the viscosity of the solution, and $D_P=2R_P$ is the diameters of the trapped object.
 
 Fig.\ref{fig4}(b,c) present the experimental results of trap stiffness when the size of trapped particles changes from 2 $\mu$m to 4 $\mu$m (light power is 20 mW). The intersection of colored dash line and solid line [Fig.\ref{fig4}(b)] represents the corner frequency. Clearly, trap stiffness increases with larger trapped Mie particles. In theory, since the size of experimental polystyrene bead is larger than wavelength, the full-wave generalized Lorenz-Mie theory and Maxwell stress tensor technique \cite{2021BesselCAB_FuxiLu,RN324,RN325} are used to calculate the trapping force and stiffness.. As shown in Fig.\ref{fig4}(c), the theoretical results are consistent with our experimental results.  Trap stiffness still  becomes larger when the size of trapped Mie particles increase.
	
	\section{Conclusion}
	In summary, we conducted a systematic investigation into the tunable autofocusing propagation and trapping performance of CSBs by tuning the beam parameters. Our experimental results demonstrate that CSBs can trap well both  Rayleigh particles and Mie particles. This work contributes to the development of optical trapping of autofoucsing beams providing new photonic tools for optical tweezers and manipulation.
	
	\begin{acknowledgments}
		This work was supported by the National Natural Science Foundation of China (11604058), the Guangxi Natural Science Foundation (2020GXNSFAA297041).
	\end{acknowledgments}
%

\end{document}